\documentclass[10pt]{article}

\usepackage{mathptmx}
\usepackage{amsmath}
\usepackage{caption}
\usepackage{subcaption}
\usepackage{graphicx}
\usepackage{lineno}
\usepackage{geometry} 
\usepackage{cite}

\usepackage[numbers,sort&compress]{natbib}

\usepackage{authblk}

\title{Spatial hole burning in thin-disk lasers and twisted-mode operation}

\author[1,2,*] {Karsten Schuhmann}
\author[1,2] {Klaus Kirch}
\author[3] {Francois Nez}
\author[4,5] {Randolf Pohl}
\author[1] {Gunther Wichmann}
\author[1,2] {Aldo Antognini}

\affil[1]{Institute for Particle Physics and Astrophysics, ETH, 8093 Zurich, Switzerland}
\affil[2]{Paul Scherrer Institute, 5232 Villigen PSI, Switzerland}
\affil[3]{Laboratoire Kastler Brossel, Sorbonne Université, CNRS, ENS-PSL University, Collège de France 4 place Jussieu, 75005 Paris, France}
\affil[4]{Johannes Gutenberg-Universit\"at Mainz, QUANTUM, Institut f\"ur Physik \& Exzellenzcluster PRISMA, 55128 Mainz, Germany}
\affil[5]{Max-Planck-Institute for Quantum Optics, 85748 Garching, Germany}
\affil[*]{Corresponding author: skarsten@phys.ethz.ch}
\begin{document}


\maketitle


\begin{abstract}
\noindent Spatial hole burning prevents single-frequency operation of thin-disk lasers when the thin disk is used as a folding mirror.
We present an evaluation of the saturation effects in the disk for disks acting as end-mirrors and as folding-mirrors explaining one of the main obstacles towards single-frequency operation.
It is shown that a twisted-mode scheme based on a multi-order quarter-wave plate combined with a polarizer provides an almost complete suppression of spatial hole burning and creates an additional wavelength selectivity that enforces efficient single-frequency operation.
\end{abstract}

\section{Introduction}

The various longitudinal modes of a laser cavity undergoing optical
amplification in the same active medium experience the phenomenon of
gain competition which leads to the prevailing of one mode
over all the others and thus to single-frequency operation~\cite{siegman1986lasers}.
A tiny difference of the effective gain  seen by a mode
is sufficient to initiate this prevailing process.
However, the domination of one mode over all the others can be
sustained only when the stimulated emission induced by this mode
causes gain saturation not only for itself (self saturation), but with
equal strength also for the other modes (cross saturation).
This cross saturation occurs when the amplification of the various
modes addresses the same regions of the active medium, so that the
amplification of one mode depletes also the population inversion
experienced by the other modes.
If the various modes address the distribution of the population inversion in different ways, as in the case of ``spatial hole burning'' (SHB) ~\cite{siegman1986lasers, Mossakowska93}, the amplification process of one mode only partially affects the population inversion experienced by the other modes.
Several modes can thus coexist in steady-state conditions in a linear laser cavity preventing single-frequency operation.

\begin{figure}[tp]
\centering
\includegraphics[width = .5\linewidth]{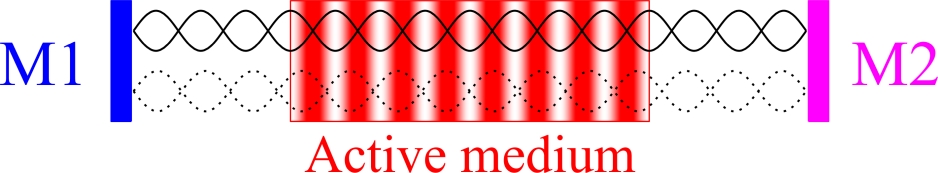}
\caption{{Scheme illustrating the principle of SHB in the active medium of a linear laser cavity. The end-mirrors (M1 and M2) are given in blue and magenta, respectively. The continuous black curves show the envelope of the time varying amplitude of a cavity mode saturating the active medium. The red color represents the position-dependent population inversion in the active medium caused by gain saturation of this mode. The dashed black curves show the envelope of one neighboring mode. This mode experiences additional gain as it addresses portions of the active medium that are not strongly saturated by the first mode. Thus, both modes can coexist in the cavity.}}
\label{fig:hole_burning_intro}
\end{figure}

The principle of SHB and its relation to single
frequency operation is illustrated in
Fig.~\ref{fig:hole_burning_intro}.
In the center of a linear cavity, the standing wave pattern of the 14$^{\mathrm{th}}$ longitudinal mode (black continuous curves) has maxima where the 13$^{\mathrm{th}}$ longitudinal mode (black dashed curves) has minima.
Therefore, the depletion of the population inversion caused by the 14$^{\mathrm{th}}$ mode reduces the gain of the 14$^{\mathrm{th}}$ mode (self saturation) but does not significantly decrease the gain of the 13$^{\mathrm{th}}$ mode (negligible cross saturation).
As a consequence, the gain competition between these two adjacent
longitudinal modes is strongly suppressed and the laser operates
simultaneously in several longitudinal modes.

\begin{figure}[bp]
\centering
\includegraphics[width= 0.4 \linewidth]{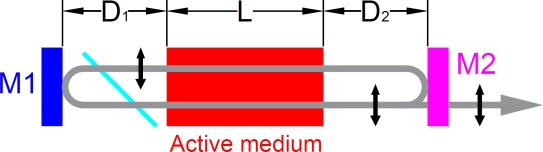}
\caption{Scheme of a linear cavity with active medium (red),
  end-mirrors (M1 and M2) (blue and magenta) and polarizer (cyan). The black
  arrows indicate the beam polarization.}
\label{fig:oscillator_geometry_simple}
\end{figure}

%
%
Spatial hole burning is often eliminated by the use of unidirectional ring-laser cavities. 
However, this is not an option for high-power thin-disk lasers 
as Faraday rotators, required to enforce the unidirectional operation, lack the power handling capability \cite{LATJ:LATJ201600017}. 
A well established method to eliminate SHB in a linear cavity, is given by the so called twisted-mode scheme \cite{Evtuhov1965, Kimura71, Smith1972, Park1984,siegman1986lasers} where the counter-propagating beams in the active medium have orthogonal polarizations so that no interference occurs.
To our knowledge, this method has so far not been applied to thin-disk lasers, except in our previous work~\cite{Aldo09}.

By reducing the length of a laser cavity (in the micrometer range
for Yb:YAG as active medium), the effect of spatial hole burning in preventing
single-frequency operation can be eliminated.
Indeed for such a short cavity only one longitudinal mode falls into the spectral region where the active medium has a gain.
Yet, these short cavities are not apt for high-power lasers as they can not sustain modes with large transverse size.
Wavelength selective elements as gratings or Fabry-Perot etalons
could be used to favor one of the modes, but also in these cases an
efficient selectivity can be achieved only for relatively short
cavity lengths.

Another method to suppress the disrupting effects of SHB and to achieve single-frequency operation is obtained by using a short active medium placed close to a cavity end-mirror.
This is often called a ``gain-at-the-end'' configuration.
In fact, as can be deduced from Fig.~\ref{fig:hole_burning_intro} in this region the intensity maxima of neighboring modes are located at similar positions so that these modes sample the active medium with nearly identical patterns.
Thus a strong cross saturation between neighboring modes exists which leads to the prevailing of the dominant mode over the others.
Because of this, a thin-disk laser with the disk acting as end-mirror M1 is suitable for high-power lasers in single-frequency operation given the mitigation of the effects of SHB and the power scalability of the thin-disk~\cite{Giesen2005,Giesen2007,Aldo09}.


{Improper understanding of the ``gain-at-the-end'' configuration has sometimes lead to erroneously neglecting SHB when the thin-disk is used as a folding mirror of the resonator.}
%
%
In such a case, the SHB pattern is more complex than
presented in Fig.~\ref{fig:hole_burning_intro} because it results from
the interference of four circulating beams as recognized by Vorholt and Wittrock in reference~\cite{Vorholt2015}.
However, their model does not include the dependence on the distance between the active medium and the end-mirror (see $D_3$ in Fig.~\ref{fig:V-shaped_amplifier}).
Spatial hole burning for disks acting as folding-mirrors has also been
discussed for mode-locked lasers \cite{Paschotta2001, Palmer2008}, though this is not applicable for cw and ns pulses.
This study aims to clarify the SHB effects taking place in thin-disk lasers and their relation to single-frequency operation.

In Sec.~\ref{Sec: The principle of spatial hole burning} we introduce the fundamentals of SHB for the simple situation presented in Fig.~\ref{fig:hole_burning_intro} similar to the studies in references~\cite{Tang1963, Roess1966, Baer1986}.
The obtained results are then applied in Sec.~\ref{Sec: thin disk as cavity end-mirror} to the case of thin-disk lasers with the
thin-disk acting as cavity end-mirror. 
In this design SHB favors longitudinal modes with a large frequency spacing from the lasing (saturating) mode, that can be easily suppressed using standard frequency selective elements. 
%
%
Section~\ref{Sec: thin-disk as bending mirror} presents SHB when the thin disk acts as a folding-mirror. 
The resulting interference of four beams in the active medium increases the susceptibility for SHB. 
In Sec.~\ref{Sec: Twisted mode thin-disk laser} we apply the twisted-mode scheme \cite{Evtuhov1965,Kimura71,Smith1972,Park1984} to this cavity layout to reduce the four-beam to a two-beam interference nearly eliminating the SHB.
It is concluded with Sec.~\ref{Sec: Q-switched thin-disk laser} presenting an implementation of the twisted-mode scheme in a Q-switched thin-disk laser \cite{Schuhmann2013}.

\begin{figure}[htbp]
\centering
\includegraphics[width=0.385\linewidth]{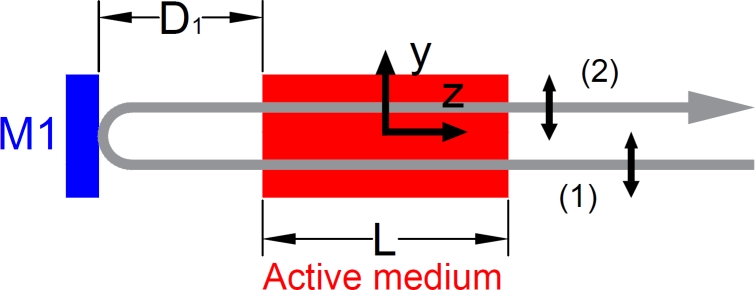}
\caption{Scheme of a dual-pass amplifier used to evaluate the SHB also for the linear cavity in Fig.~\ref{fig:oscillator_geometry_simple}. Same color convention is used as in Fig.~\ref{fig:oscillator_geometry_simple}. Due to the absence of resonant  condition the possible values of $\lambda$ and $\lambda_0$ are not constrained.}
\label{image1_b}
\end{figure}

\section{Spatial hole burning for a linear cavity} 
\label{Sec: The principle of spatial hole burning}

In this section, the well known principle of SHB is derived for the linear cavity in Fig.~\ref{fig:oscillator_geometry_simple} by using the generic building block of an amplifier given in Fig.~\ref{image1_b}. In this amplifier a saturating Gaussian beam at a wavelength $\lambda_0$ passes an active medium, travels to a mirror M1 that is reflecting the beam back onto itself. 
Thus the beam crosses the medium a second time. 
The origin of the coordinate system used to compute SHB is located in the center of the active medium with $z$ the propagation and $x$,$y$ the transverse directions.
The position- and time-dependent amplitude of the forward propagating beam inside the active medium  ($z\in[-L/2, +L/2]$ with the length of the active medium $L$) can be written as:
\begin{equation}
  E_1(x,y,z,t, \lambda_0) = E_0 \cos{\left(2\pi~\left(-\frac{n~z}{\lambda_0}- \nu_0~t\right)\right)} \cdot \mathrm{exp}\left(-\frac{x^2 + y^2}{2 w^2}\right),
\end{equation}
with \textit{w} the width of the transverse profile of the beam, \textit{n} the refractive index of the active medium, $t$ the time, $E_0$ the amplitude of the beam on its symmetry axis, $\nu_0$ and $\lambda_0 = c/\nu_0$ the frequency and wavelength of the beam, respectively.
For simplicity we neglected the increase of amplitude occurring in the active medium. This simplification allows for analytical results and is well justified for low-gain media as the thin-disk.  
Furthermore it provides qualitatively valid results also for high-gain media.
Note that the laser gain will be accounted for separately (see e.g. Eq.~(\ref{eq:gain_int})).
The corresponding backward propagating beam is given by
\begin{equation}
   E_2(x,y,z,t, \lambda_0) = E_0 \cos{\left(2\pi~\left(\frac{2d+n~z}{\lambda_0} - \nu_0~t\right)\right)} \cdot \mathrm{exp}\left(-\frac{x^2 + y^2}{{2~w}^2}\right),
\end{equation} 
where $d = D_1 + n L/2$ is the optical distance between the center of the active medium and M1 with $D_1$ the distance between the active medium and M1 (cf. Figs.~\ref{fig:hole_burning_intro}~and~\ref{image1_b}).
The total amplitude in the active medium resulting from the superposition of the two beams is thus
\begin{equation}
     E(x,y,z,t, \lambda_0) = E_1 + E_2=2 E_0 \cos{\left(2\pi~\left( \frac{d}{\lambda_0}-\nu_0 t\right)\right)}\cos{\left(2\pi\frac{d + n z}{\lambda_0 }\right)} \cdot \mathrm{exp}\left(-\frac{x^2 +y^2}{2~w^2}\right).
  \label{eq:interference-simple}
\end{equation}
The position dependency of the time-averaged intensity distribution resulting from the two counter-propagating beams is given by
\begin{equation}
  I\left(x,y,z,\lambda_0 \right) = 2 I_0 \left(1+\cos{\left(\mathrm{4}\pi~\frac{n~z+d}{\lambda_0 }\right)}\right) \mathrm{exp}\left(-\frac{x^2 + y^2}{w^2}\right),
  \label{eq:intensity}
\end{equation}
where $I_0 = c~n\varepsilon_0|E_0^2|/2$ is the intensity of a single laser beam on its symmetry axis.
From this intensity distribution $I(x,y,z,\lambda_0)$, it is possible to obtain the position-dependent density of the population inversion in the active material $N(x,y,z, \lambda_0)$ by using the known rate equation for a four-level laser~\cite{siegman1986lasers}

\begin{equation}
 \frac{dN (x,y,z,t, \lambda_0)}{dt} = R_p - N(x,y,z,t, \lambda_0)\frac{I(x,y,z,t,\lambda_0 )}{F_\mathrm{sat}} - \frac{N(x,y,z,t, \lambda_0 )}{\tau},
\end{equation}
where $R_p$ is the pump rate (homogeneous within
the active medium), $F_\mathrm{sat}$ the saturation fluence
and $\tau $ the lifetime of the upper state.
The decrease of the upper state population by the stimulated emission is proportional to the local intensity of the circulating light $I(x,y,z,\lambda_0 )$.
The steady-state solution of this rate equation ($dN/dt = 0$) reads
\begin{eqnarray}
  N(x,y,z, \lambda_0) &=&\frac{R_p \tau }{1+ \tau \frac{I(x,y,z,\lambda_0 )}{F_\mathrm{sat}} } \\
  &\approx & R_p \tau - R_p \tau^2 \frac{I(x,y,z,\lambda_0 )}{F_\mathrm{sat}} + \mathcal{O}\left({I^2}\right).
  \label{linearized_rate_EQ}
\end{eqnarray}
In Eq.~(\ref{linearized_rate_EQ}) we assumed that $\tau \frac{I(x,y,z,\lambda_0 )}{F_\mathrm{sat}}\ll 1$, which is valid for a weakly saturating beam intensity
\footnote{For efficient laser operation high saturation of the active medium is required. In this case this assumption does not hold. Nevertheless, the general behaviour presented in this study remains valid.}, e.g. for a laser cavity with an out-coupling mirror of low reflectivity.
The position dependent gain $g\left(x,y,z, \lambda_0\right)$ is
proportional to the local density of the population inversion and can be expressed as
\begin{equation}
   g\left(x,y,z, \lambda_0\right)\approx g_0\mathrm{-}g_0\tau
  \frac{I\left(x,y,z,\lambda_0 \right)}{F_\mathrm{sat}},
   \label{eq:gain}
\end{equation}
where $g_0$ is the unsaturated gain.

So far we described the SHB effect of the dominant laser beam with a wavelength $\lambda_0$ amplified in a dual-pass configuration as depicted in Fig.~\ref{image1_b}.
As a next step we evaluate the saturation spectrum of the active medium experienced by a second (probing) beam at a wavelength $\lambda$.
This beam forms a similar interference pattern $I(x,y,z,\lambda)$ in the active medium slightly shifted relative to $I(x,y,z,\lambda_0)$. 
The cross-saturation gain $G$ experienced by this beam is given by the integral 
\begin{equation}
  G(\lambda)~=~\frac{\int \int \int I(x,y,z,\lambda )~g(x,y,z,\lambda_0)~dx~dy~dz}{P_\mathrm{laser}}.
  \label{eq:gain_int}
\end{equation}
where $P_\mathrm{laser}$ is the incident laser power at wavelength $\lambda_0$.
The gain $G$ can be described as the product of the unsaturated laser gain $G_0$ and the saturation spectrum $S_\mathrm{SHB}(\lambda)$ so that
\begin{equation}
   P_\mathrm{out} =e^{G(\lambda)}~P_\mathrm{in} = e^{S_\mathrm{SHB}(\lambda)~G_0(\lambda)}~P_\mathrm{in} .
   \label{eq:Descrip_gain}
\end{equation}
The saturation spectrum $S_\mathrm{SHB}(\lambda)$ for the dual-pass configuration of Fig.~\ref{image1_b} is thus given by
\begin{equation}
  \begin{array}{c}
  S_\mathrm{SHB}(\lambda) = \frac{G}{G_0} =\frac{\int\limits_{-\infty}^\infty \int\limits_{-\infty}^\infty \int\limits_{-L/2}^{L/2} I\left(x,y,z,\lambda \right)~g(x,y,z,\lambda_0)~dz~dx~dy}{\int\limits_{-\infty}^\infty \int\limits_{-\infty}^\infty \int\limits_{-L/2}^{L/2}I(x,y,z,\lambda)~g_0~dz~dx~dy}.
  \end{array}
  \label{eq:gain_HB}
\end{equation}
Combining Eq.~(\ref{eq:gain}) with Eq.~(\ref{eq:gain_HB}) we find
\begin{equation} 
  \begin{array}{cc}
   S_\mathrm{SHB}(\lambda)=\\  1-\frac{\tau}{I_0  F_\mathrm{sat} L}\int\limits_{-\infty}^\infty \int\limits_{-\infty}^\infty \int\limits_{-L/2}^{L/2} I\left(x,y,z,\lambda _0\right)~I\left(x,y,z,\lambda \right)~dz~dx~dy~.
  \end{array}
  \label{eq:gain_HB_almost_final}
\end{equation}
\begin{figure}[tp]
\centering
\includegraphics[width=.5\linewidth]{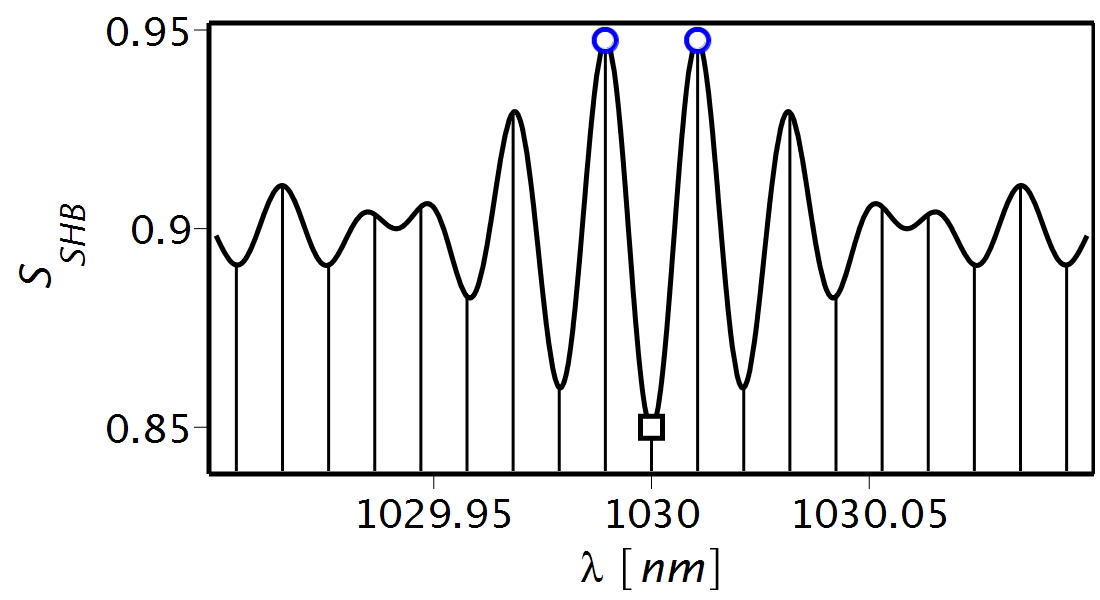}
\caption{Saturation spectrum $S_\mathrm{SHB}$ of the amplifier given in Fig.~\ref{image1_b} computed with Eq.~\ref{eq:gain_HB_final} assuming $D_1=40$~mm, $L=10$~mm, \textit{$ 2\tau I_{0} / F_\mathrm{sat}~=$}~10\% and the saturating wavelength $\lambda_{0}=1030$~nm.
The plot also represents $S_\mathrm{SHB}$ of the linear cavity in Fig.~\ref{fig:oscillator_geometry_simple} with $D_1 = D_2 = 40$~mm.
The vertical lines indicate the wavelength of the cavity eigenmodes.
The gain experienced by the saturating mode is indicated by the black empty square, the gain experienced by its neighboring modes by the empty blue circles. }
\label{fig:gain_SHB}
\end{figure}

\noindent Inserting the time-averaged intensity of Eq.~(\ref{eq:intensity}) into Eq.~(\ref{eq:gain_HB_almost_final}), and assuming that $d$ and $L$ are much larger than $\lambda $ and $\lambda_0$, and also that $\Delta \equiv (\lambda_{0} - \lambda )/( \lambda_{0} \lambda )\ll 1 /\lambda_{0}$ we find
\footnote{To obtain this result we have made use of the approximation
\begin{equation}
\int\limits_{-L/2}^{L/2} \left(1+ \cos \left(2\pi \frac{n~z+ d}{\lambda_0}  \right)\right)\left(1+ \cos \left(2\pi \frac{n~z+d}{\lambda}\right)\right)dz \approx L\left(1+\frac{1}{2} \mathrm{sinc}\left(\pi~n~L~\Delta \right)~\cos \left(2\pi~d~\Delta \right) \right).
\end{equation}
} 
\begin{equation}
  S_\mathrm{SHB}(\lambda )~\approx~1-2 \tau \frac{I_0}{F_\mathrm{sat}}\left(1+\frac{1}{2} \mathrm{sinc}(\pi~n~L~\Delta)~\cos (2\pi~d~\Delta)\right).
  \label{eq:gain_HB_final}
\end{equation}
A plot of the saturation spectrum $S_\mathrm{SHB}(\lambda )$ for realistic values of $d$ and $L$ is shown in Fig.~\ref{fig:gain_SHB} as a function of the wavelength $\lambda$ of the probe beam assuming a saturating beam at $\lambda_0 = 1030$~nm.
When applying this results to a cavity, it is necessary to take into account that the wavelength $\lambda$ and $\lambda_0$ can only have discrete values as a consequence of the resonance conditions.
For the simplified cavity layout in
Fig.~\ref{fig:oscillator_geometry_simple} with $D_1 = D_2$, the longitudinal modes neighboring the saturating mode are located exactly at the two maxima of $S_\mathrm{SHB}(\lambda )$ as indicated by the blue hollow circles in Fig.~\ref{fig:gain_SHB}.
These two modes experience thus a gain decrease due to saturation
effects significantly smaller compared to the gain decrease at the
saturating wavelength $\lambda_0$.
Therefore amplification of these neighboring modes occurs leading
to disruption of single-frequency operation.
Because the spectral distance between modes for realistic layouts of rod lasers with typical cavity round-trip length of about 0.1~m is 1.5~GHz (5~pm at 1030~nm), the suppression of the neighboring modes using selective elements would require extremely narrow spectral filters with active regulation to prevent significant cavity losses at the operating wavelength $\lambda_0$.

\section{Spatial hole burning for a thin disk acting as cavity end-mirror}
\label{Sec: thin disk as cavity end-mirror}

\begin{figure}[bp]
\centering
\includegraphics[width=0.375 \linewidth]{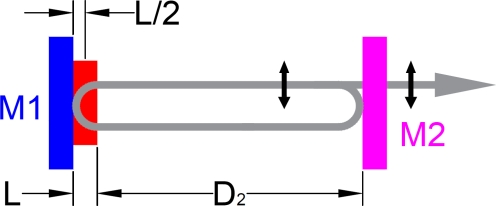}\\[3mm]
\caption{Scheme of a thin-disk cavity where the disk is acting also as end-mirror M1. Same color code as in Fig.~\ref{fig:oscillator_geometry_simple}.}
\label{fig:thin-disk_end-mirror_scheme}
\end{figure}

\begin{figure}[tp]
\centering
\includegraphics[width=0.4\linewidth]{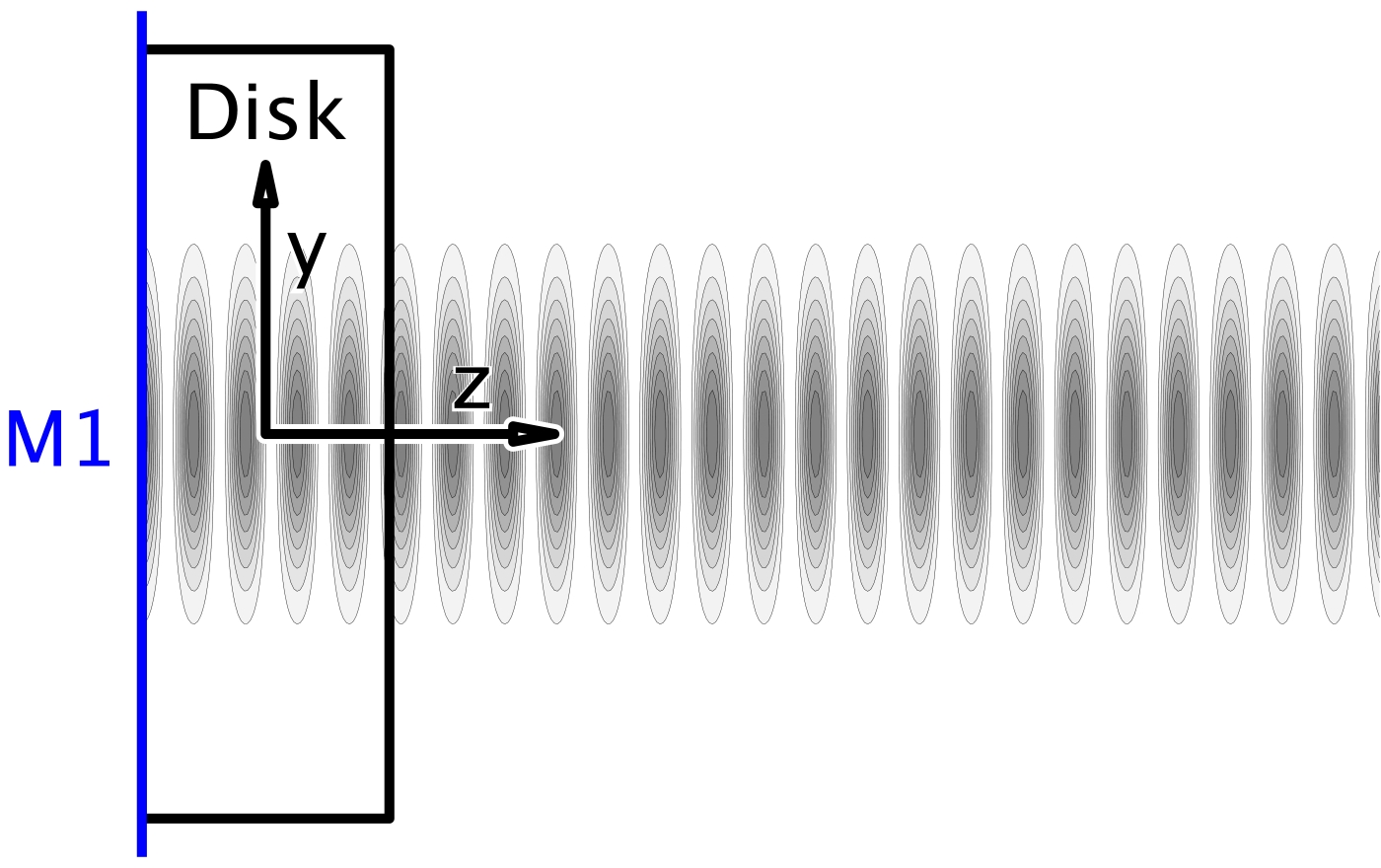}
\caption{Intensity pattern in the disk when the back side of the disk is reflecting the beam in itself. Due to the two-beam interference, the maximal intensity is $4 I_0$. The boundary conditions at mirror M1 impose the interference pattern to have an anti-node at the M1 position. Therefore the positions of the maxima and minima of the interference pattern change slowly with the wavelength $\lambda$. }
\label{fig:thin-disk_end-mirror_interference}
\end{figure}

For a thin disk acting as a cavity end-mirror M1 (see  Fig.~\ref{fig:thin-disk_end-mirror_scheme}), the effects related with SHB are mitigated.
Indeed  for this ``gain-at-the-end'' configuration the intensity pattern in the active medium (see Fig.~\ref{fig:thin-disk_end-mirror_interference}) change so slowly with the wavelength $\lambda$ that the  maxima and minimima of the intensity pattern of two neighboring axial cavity modes are located at similar positions (cf. Fig.~\ref{fig:hole_burning_intro}).
This implies that the cross saturation between neighboring modes is nearly identical to the self saturation. 
In other words, the saturation spectrum $S_\mathrm{SHB}(\lambda)$ is a slowly varying function of $\lambda$.
%

The reduction of gain caused by the saturation spectrum $S_\mathrm{SHB}(\lambda)$ for thin-disk lasers with disks acting as cavity end-mirrors can be quantified using Eq.~(\ref{eq:gain_HB_final}) and setting $D_1 = 0$ (cf. Figs.~\ref{fig:oscillator_geometry_simple} with \ref{fig:thin-disk_end-mirror_scheme})
\begin{equation}
S_\mathrm{SHB}(\lambda ) \approx  1 - 2 \frac{\tau I_0}{F_\mathrm{sat}}\left(1+\frac{1}{2}  \mathrm{sinc} (\pi~n~L~\Delta )\right)\ .
\label{eq:end-mirror}
\end{equation}
\begin{figure}[bp]
\centering
  \begin{subfigure}[t]{ .47\linewidth}
    \centering
    \includegraphics[width=\linewidth]{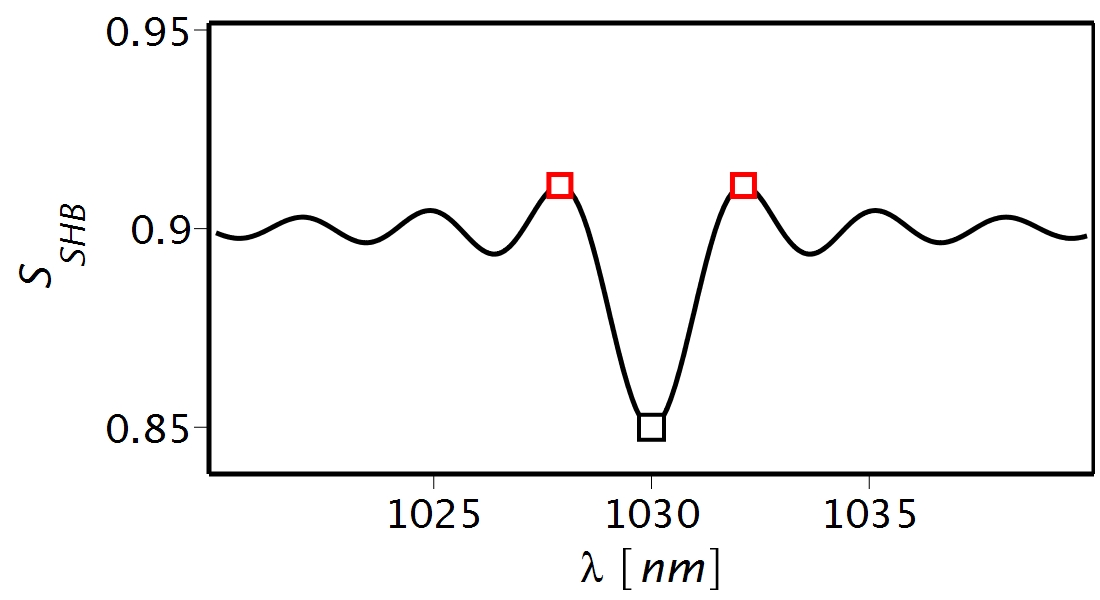} 
    \caption{} \label{Wide}
  \end{subfigure}
  \begin{subfigure}[t]{ .47\linewidth}
    \centering
    \includegraphics[width=\linewidth]{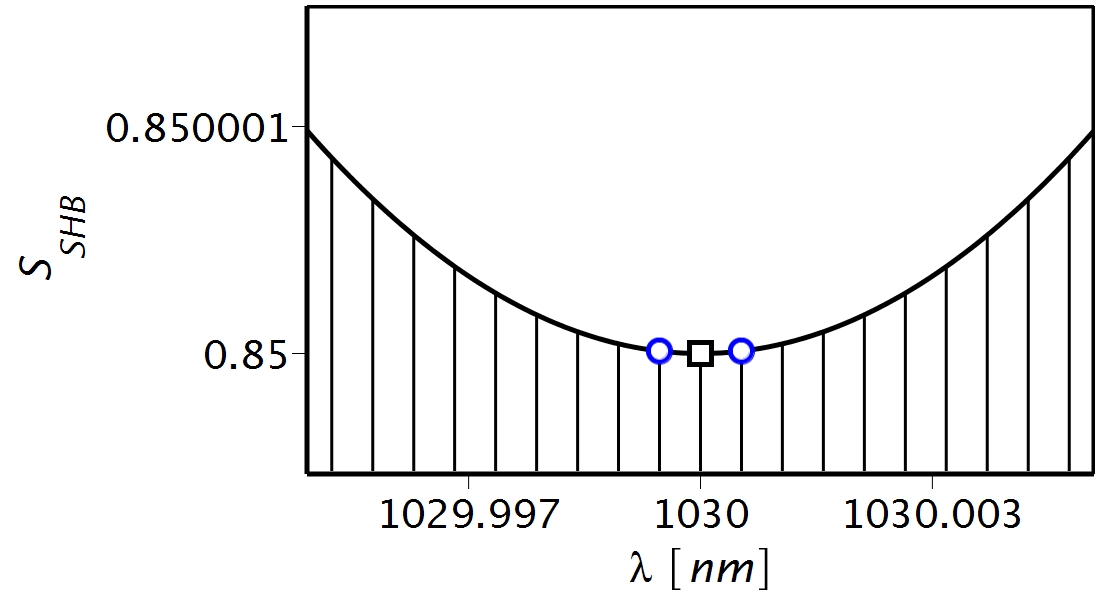} 
    \caption{} \label{Narrow}
  \end{subfigure}
  \caption{(a) Saturation spectrum $S_\mathrm{SHB}$ of a thin-disk caused by a beam reflected at the disk rear side as a function of the wavelength $\lambda$ calculated using Eq.~(\ref{eq:end-mirror}).
    The saturating mode is at $\lambda_{0} = 1030$~nm, $D_1 = 0$~m, $L = 0.4$~mm, $w=1$~mm and \textit{$2 \tau I_{0} / F_\mathrm{sat} =$}~10\%. 
    The self-saturation is indicated by the black empty square. The red empty squares indicate the spectral position of minimal cross-saturation.
    (b) Zoom by a factor of $2\times 10^4$ of the plot in panel (a) around the saturating wavelength. 
    The vertical lines indicate the wavelengths of the eigenmodes for the cavity of Fig.~\ref{fig:thin-disk_end-mirror_scheme} assuming $D_2 = 2$~m.
The cross saturation of the neighboring modes which is indicated by the empty blue circles is nearly identical to the self saturation.}
    \label{fig:gain-HB_end-mirror}
\end{figure}
$S_\mathrm{SHB}$ is plotted in Fig. \ref{fig:gain-HB_end-mirror} as a function of the wavelength $\lambda$ for a realistic thin-disk laser ($D_2=2$ m) having a saturating mode at $\lambda_0 = 1030$~nm.
A long cavity was assumed in this case to provide an intra-cavity beam with large $w$ required for the high-power operation.
Because of this, the wavelength difference between neighboring modes in the 1030~nm region is only about 0.25~pm (75~MHz) apart, as indicated by the vicinity of the blue hollow circles and the black hollow square in Fig. \ref{Narrow}.
Thus, competition between the various modes located around the broad minimum centered at 1030~nm exists, which can be used to achieve single-frequency operation.
In Fig. \ref{Wide} the red squares indicate the maxima of the saturation spectrum $S_\mathrm{SHB}$ at a spectral distance of 2.11~nm (615~GHz) from $\lambda_0$.
These modes could grow in the laser cavity and disrupt single-frequency operation.
However, due to the large spectral distance of these maxima, the laser operation at these wavelengths can be easily suppressed by introducing a standard frequency selective element (e.g. a birefringent filter or a grating~\cite{Walther1970}) without active stabilization.

%

\section{Spatial hole burning for a thin-disk as a folding-mirror}
\label{Sec: thin-disk as bending mirror}

\begin{figure}[tp]
\centering
\includegraphics[width=0.35 \linewidth]{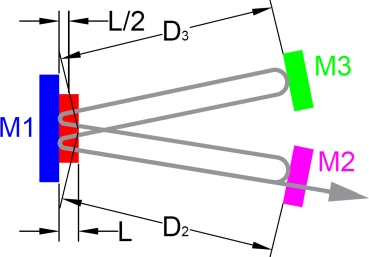}
\caption{Scheme of a thin-disk cavity with V-shaped layout where
  the disk is acting as folding-mirror M1. The active medium is given in
  red, the end-mirrors M2 and M3 in green and magenta, respectively. }
\label{fig:V-shape}
\end{figure}

A more stable output beam for variations of the thin-disk thermal lens can be obtained by using the thin disk as a folding mirror in a V-shaped cavity (see Fig.~\ref{fig:V-shape}). 
The stability properties of a laser containing a thermal lens have been studied in  \cite{Magni1986}. 
The thermal lens splits the cavity in two parts.
The part which is longer (effective length given by the B-element of the corresponding ABCD-matrix) has a complex beam parameter $q$ which exhibits a linear (first-order) dependence on variations of the thermal lens.
By contrast, the $q$-parameter in the shorter part exhibits a second-order dependence on variations of the thermal lens around the center of stability. 
Thus, to minimize variations of size and divergence of the output beam, 
the out-coupling mirror has to be placed in the short part of the cavity.
When the thin disk is used as an end-mirror of the cavity, it would be necessary to use the rear side of the disk as an out-coupling mirror, which is not feasible.
As a result the beam out-coupled  from a thin-disk laser with the disk acting as end-mirror suffers from a large sensitivity to variations of the thermal lens of the active medium.
On the contrary, in a V-shaped cavity the disk is used as folding mirror, and the out-coupling mirror can be placed in the short (stable) part of the cavity leading to an out-coupled beam insensitive to variations of the thermal lens \cite{Aldo09}.

\begin{figure}[tp]
\centering \includegraphics[width=0.3 \linewidth]{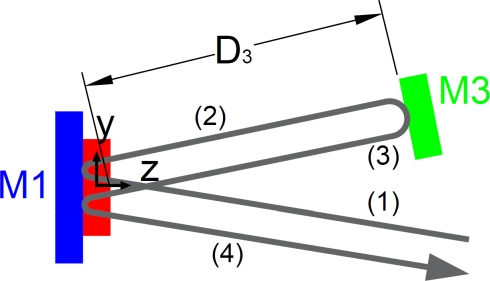}
\caption{Scheme of a 4-pass thin-disk amplifier used to evaluate the SHB
  occurring in the active medium of the thin-disk cavity of
  Fig.~\ref{fig:V-shape} with V-shaped layout. Same color code as in
  Fig.~\ref{fig:V-shape} is used. Indicated are also the reference
  system and the four beams with amplitude $E_1$, $E_2$, $E_3$ and
  $E_4$.}
\label{fig:V-shaped_amplifier}
\end{figure}

In contrast to the simple linear cavity of Fig.~\ref{fig:thin-disk_end-mirror_scheme}, in the V-shaped cavity with disk acting as folding-mirror M1, the consideration of the SHB requires the evaluation of the interference of 4 beams similar to what was presented in~\cite{Vorholt2015}.
Following the same methodology of Sec.~\ref{Sec: The principle of spatial hole burning} to compute the interference pattern we proceed first by omitting M2 and computing the interference of the four beams for the amplifier configuration shown in Fig.~\ref{fig:V-shaped_amplifier}.
In a second step, the other end-mirror (M2) is considered: it provides the resonance condition for the cavity.
%

The amplitude of the first beam $E_{1}(x,y,z,t, \lambda_0)$ traveling within the
active medium with refractive index $n$ from the disk front side
towards the disk rear side can be described as
\begin{equation}
E_1(x,y,z,t, \lambda_0)= E_0 \cdot\cos{\left(2 \pi \frac{-n z \cos{(\alpha )} + n y \sin{(\alpha )}}{\lambda_0} - 2\pi \nu_0 t\right)}\cdot \mathrm{exp}\left(-\frac{x^2 + \left( (z+L/2) \sin{(\alpha )} - y \cos{(\alpha)}\right)^2}{2~w^2}\right),
\end{equation}
where $\alpha$ is the propagation angle relative to the disk normal
within the active medium.
Again, we assume that the origin of the reference system is located at the
center of the disk, thus at a distance $L/2$ from the disk's rear and front
sides.
The 2$^{\mathrm{nd}}$ beam originates from the reflection on the high-reflection (HR) at the rear side of the disk (M1).
Within the active medium the amplitude of this second beam $E_2
(x,y,z,t, \lambda_0)$ is given by
\begin{align}
E_2 (x,y,z,t, \lambda_0) = E_0 & \cdot\cos{\left(2\pi~\frac{( L + z) n \cos{(\alpha )} + n y \sin{(\alpha)}}{\lambda_0} - 2\pi \nu_0 t \right)} \label{eq:second-beam} \\ 
& \cdot \mathrm{exp}\left(-\frac{x^2 + {\left( (z + L/2) \sin{(\alpha)} + y \cos{(\alpha)}\right)}^2}{2 w^2}\right). \nonumber
\end{align}
A phase delay and an inversion of the propagation in $z$-direction has been used to obtain this second beam from the first one, while size and direction in $x$- and
$y$-directions are unchanged.
This beam travels to M3 (given in green in Fig.~\ref{fig:V-shaped_amplifier}) that reflects the beam in itself (inversion of all three propagation directions) back towards the active medium.
The amplitude of this back-reflected beam traveling in the active medium from the front-side to the rear-side takes the form
\begin{align}
E_3 (x,y,z,t, \lambda_0) = E_0 & \cdot \cos{\left( 2\pi~\frac{2 d + n (L-z) \cos{(\alpha )} - n y \sin{(\alpha )}}{\lambda_0} - 2\pi \nu_0 t \right)} \label{eq:third-beam}\\
& \cdot \mathrm{exp}\left(-\frac{ x^2 +{\left( (z + L/2) sin (\alpha) - y \cos{(\alpha)}\right)}^2}{2 w^2}\right), \nonumber
\end{align}
where $d=D_3+nL/2$ is the optical distance between the active medium center and M3, with $D_3$ and $L$ as defined in Fig.~\ref{fig:V-shaped_amplifier}.

After a second reflection on the HR-coating of the disk, a fourth beam moving from the disk rear-side to the front-side is ensued whose amplitude takes the form
\begin{align}
E_4(x,y,z,t,\lambda_0) = E_0 &\cdot\cos{\left(2\pi\frac{2d + (2 L + z)n \cos{(\alpha)} - n  y \sin{(\alpha)}}{\lambda_0} - 2\pi \nu_0 t \right)} \label{eq:forth-beam} \\
&\cdot \mathrm{exp}\left(-\frac{x^2 + {\left( (z + L/2)  \sin{(\alpha)} + y \cos{(\alpha)}\right)}^2}{2~w^2}\right)\ .  \nonumber    
\end{align}
\begin{figure}[tp]
\centering
\includegraphics[width=.4\linewidth]{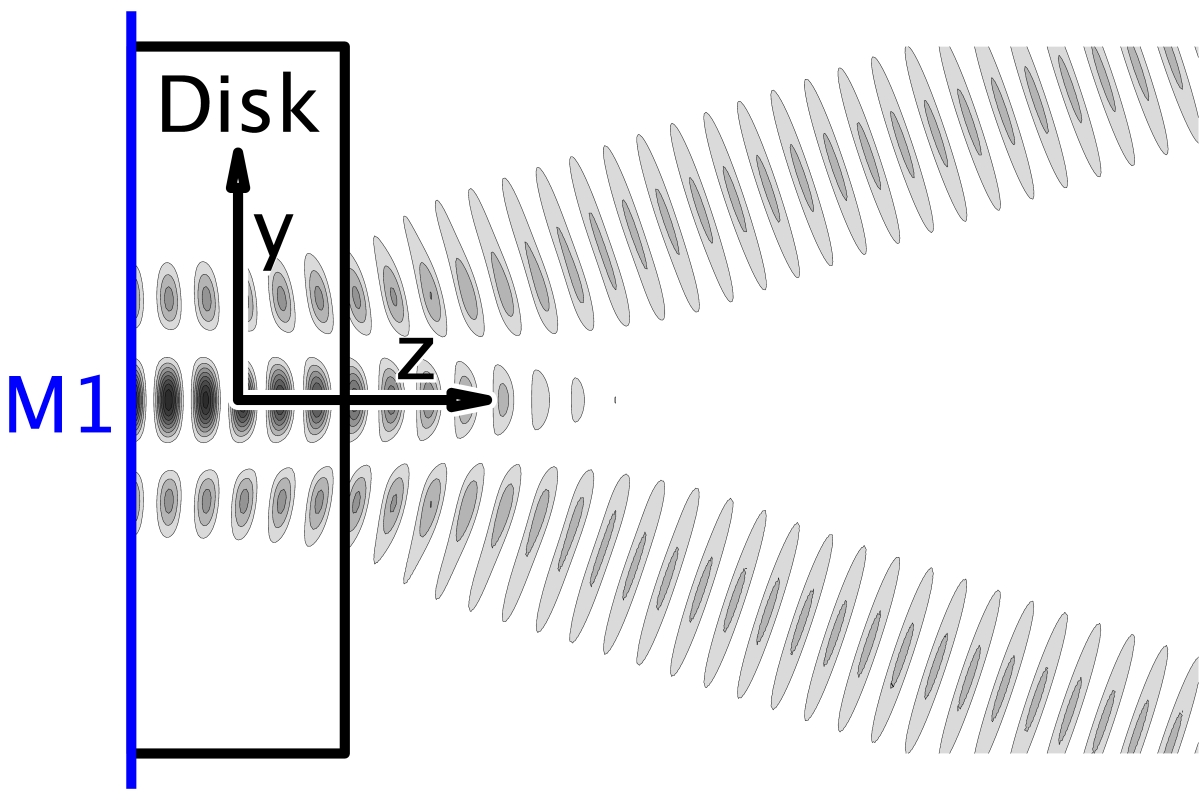}
\caption{ Intensity pattern in the disk for a disk acting as a folding-mirror in a V-shaped cavity. Due to the four-beam interference, the peak intensity is $16~I_0$. This pattern results from the interference of four Gaussian beams in the configuration of Fig.~\ref{fig:V-shaped_amplifier}. Because of the non-normal incidence within the thin disk longitudinal and transversal interference patterns appear as described by Eq.~(\ref{Eq:intensity 4 wave}).}
\label{fig:four_beams_interference}
\end{figure}
The time-averaged intensity distribution resulting from the interference of the four traveling waves takes the form

\begin{align}
I(x,y,z, \lambda_0 ) = 4 I_0 \left(1 + \cos{\left(4 \pi \frac{n z \cos{(\alpha)} + L/2}{\lambda_0}\right)}\right)  \cdot \left(1 + \cos{\left(4\pi\frac{n~y~\sin{(\alpha)}+d}{\lambda_0}\right)}\right) \mathrm{exp}\left(-\frac{x^2 + y^2}{w^2}\right)
\label{Eq:intensity 4 wave}
 \end{align}
when walk-off effects are neglected. This approximation is justified given the small thickness of the disk ($L<0.5$~mm) relative to the typical size of the cavity eigenmodes ($w>1$~mm) and the small angle ($\alpha<5^{\circ}$).

An example of such  an intensity pattern is shown in Fig.~\ref{fig:four_beams_interference}.
By inserting Eq.~(\ref{Eq:intensity 4 wave}) into Eqs.~(\ref{eq:gain}) and (\ref{eq:gain_HB}) the gain decrease related with the saturation effects is obtained:
\begin{equation}
S_\mathrm{SHB}(\lambda) \approx 1 - 4 \frac{\tau I_0}{F_\mathrm{sat}} \left(1 + \frac{1}{2} \mathrm{sinc}\left(\pi~n~L~\Delta~\cos (\alpha)\right)\right) \cdot \left(1 + \frac{1}{2} \cos \left( 4\,\pi\,\Delta\,d \right)~  \exp\left(-2 \Delta^2 \pi^2 n^2 w^2 \sin^2 \left( \alpha \right)  \right) \right).
\label{eq:G_HS_four_beams}
\end{equation}
\begin{figure}[tp]
\centering
  \begin{subfigure}[t]{.47\linewidth}
    \centering
    \includegraphics[width= \linewidth]{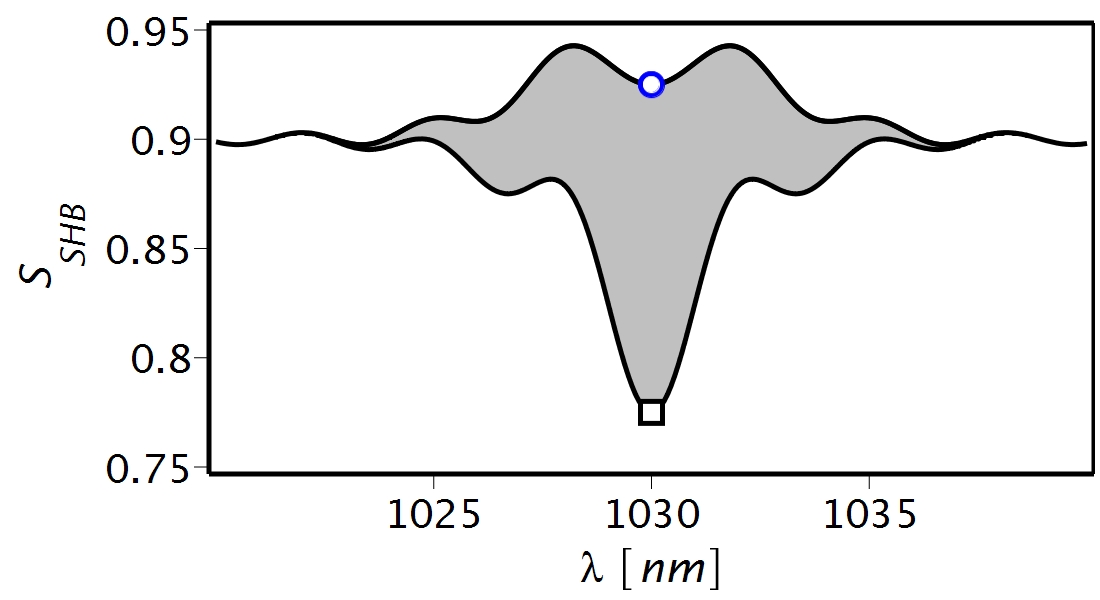} 
    \caption{} \label{Wide_Fold}
  \end{subfigure}
  \hfill
  \begin{subfigure}[t]{ .47\linewidth}
    \centering
    \includegraphics[width= \linewidth]{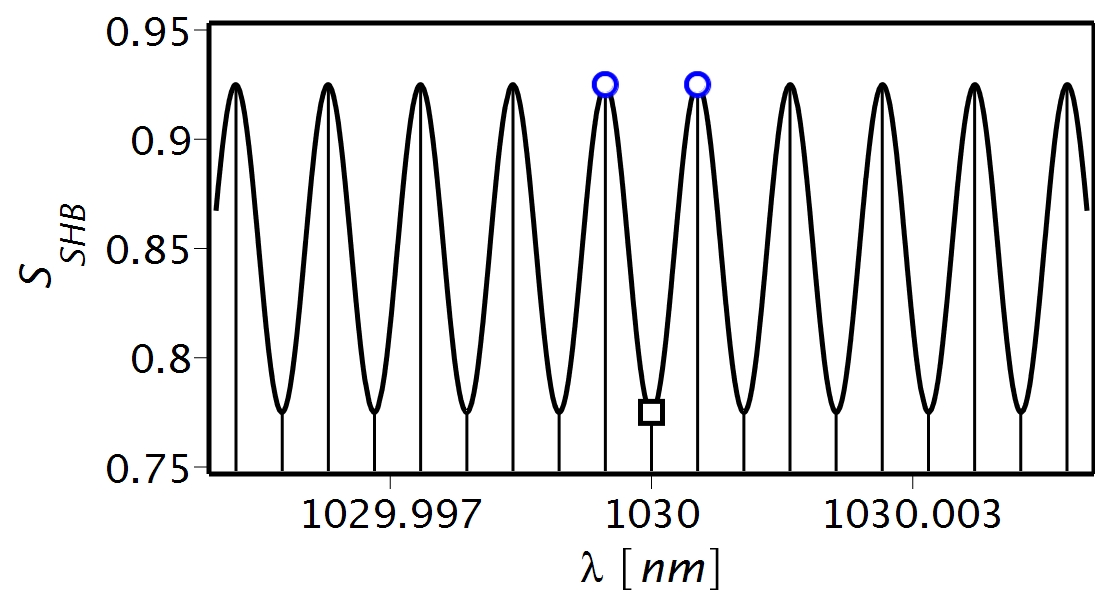} 
    \caption{} \label{Narrow_Fold}
  \end{subfigure}
\caption{(a): Saturation spectrum $S_\mathrm{SHB}(\lambda)$ of the 4-pass thin-disk amplifier from Fig.~\ref{fig:V-shaped_amplifier} calculated using Eq.~(\ref{eq:G_HS_four_beams}) and assuming a saturating beam at $\lambda_{0} = 1030$~nm, $D_3 = 1$~m, $L = 0.4$~mm, $\alpha=4.5^\circ$, $w=1$~mm and $4 \tau I_{0} / F_\mathrm{sat}~=$~10\%. 
The gain experienced by the saturating beam is indicated by the black empty square. 
The gray shaded region represents rapid (unresolved) oscillations. 
(b): Zoom in by a factor of $2\times 10^4$ around the saturating wavelength to present the the fast oscillations of the $S_\mathrm{SHB}(\lambda)$.
The vertical lines indicate the possible wavelengths of the cavity eigenmodes given in Fig.~\ref{fig:V-shape} with $D_2=D_3 = 1$~m.
}
\label{fig:G_SHB_four_waves_1}
\end{figure}
The saturation spectrum $S_\mathrm{SHB}\left(\lambda \right)$ for $D_3=1$~m and $\alpha=4.5^\circ$ is shown in Fig.~\ref{Wide_Fold} as a function of the probe beam  wavelength $\lambda$.
The self saturation is depicted by the the black square.
The gray shaded region in the spectral plot of $S_\mathrm{SHB}$ represents rapid oscillations caused by the large optical distance $D_3$ between the active medium and the end-mirror M3.
These rapid oscillations are visible in Fig.~\ref{Narrow_Fold} where the horizontal axis is magnified by a factor of $2\times 10^4$  in the wavelength region around the saturating wavelength $\lambda_0 = 1030$~nm.
In the corresponding cavity ($D_3=D_2=1$~m) the modes (indicated by the blue empty circles) nearest to the saturating mode  experience the maximal values of $S_\mathrm{SHB}$.
As the two neighboring modes (and others) have a larger small-signal gain, they will grow until they significantly contribute to the saturation of the gain medium.
Hence, single-frequency operation is disrupted.
Also in this case a suppression of these modes would require an extremely narrow-band filter with active stabilization to prevent significant losses at the wavelength of the saturating mode.

\section{Twisted-mode operation of a V-shaped thin-disk laser}
\label{Sec: Twisted mode thin-disk laser}

\begin{figure}[htbp]
\centering
  \begin{subfigure}[t]{.42 \linewidth}
    \centering
    \includegraphics[width= \linewidth]{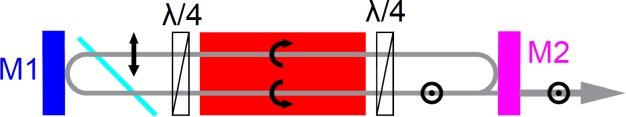} 
    \caption{} \label{fig:twisted-mode_classic}
  \end{subfigure}
  \hfill
  \begin{subfigure}[t]{ .42 \linewidth}
    \centering
    \includegraphics[width= .9\linewidth]{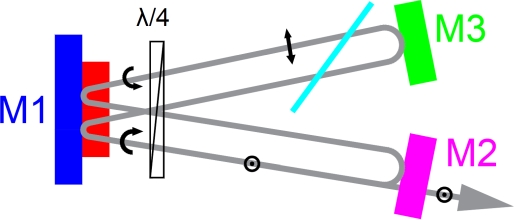} 
    \caption{} \label{fig:twisted-mode_new}
  \end{subfigure}
\centering
\caption{(a): Linear cavity with a twisted-mode scheme obtained
  by inserting two quarter-wave plates around the active medium (red). (b):
  cavity of a V-shaped thin-disk laser with a twisted-mode
  scheme obtained using only one multi-order quarter-wave
  plate. The polarization of the circulating beams is given by
  the black arrows and circles, the polarizer is in cyan.}
\end{figure}

The twisted-mode scheme~\cite{Evtuhov1965, Kimura71, Smith1972, Park1984} represents a well-known possibility to eliminate SHB.
Its principle is to insert two quarter-wave plates, one before and one after the active medium, as presented in Fig.~\ref{fig:twisted-mode_classic} such that there is no interference in the active medium, because the two counter-propagating beams have orthogonal polarizations.
The absence of the interference between the two circulating beams leads to the complete elimination of the SHB paving the way for single-frequency operation~\cite{Bollen1987, Adams1993, Pan2005, Cong2016}.

\begin{figure}[tp]
\centering
\includegraphics[width=.4 \linewidth]{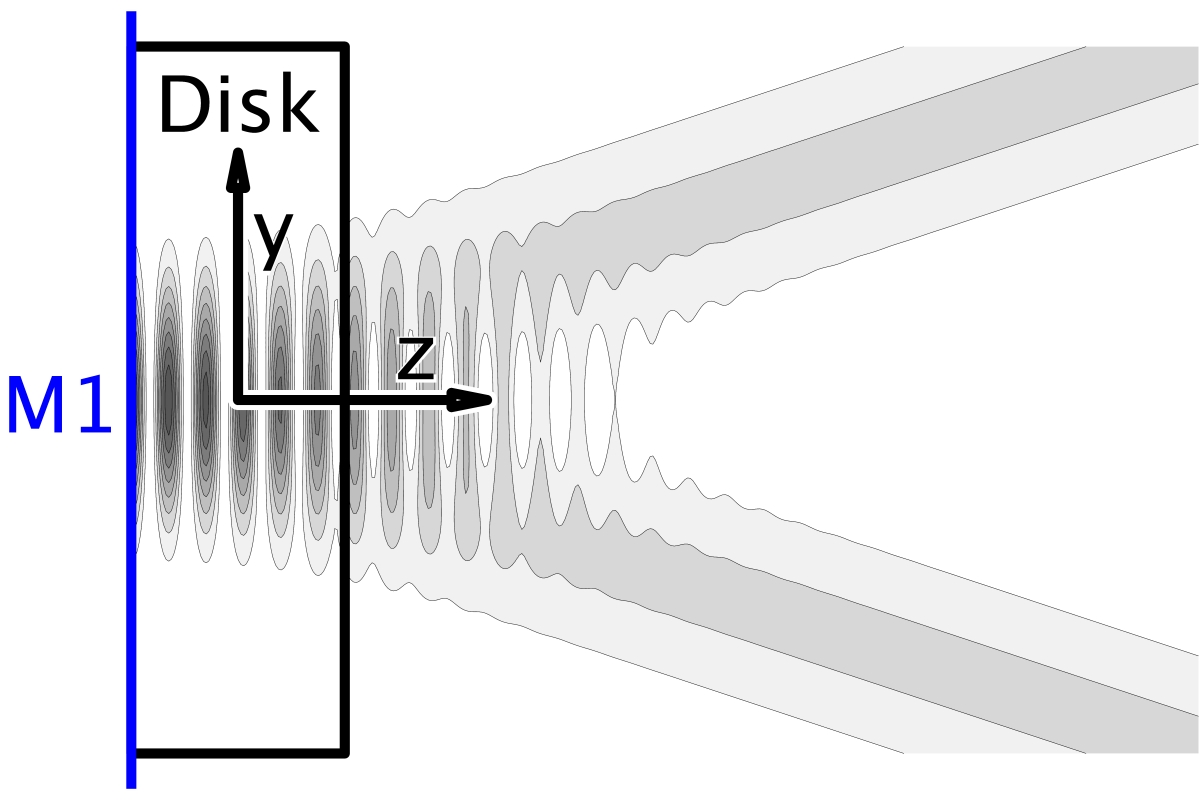}
\caption{Intensity pattern in the disk for a disk acting as a folding-mirror in a V-shaped cavity and twisted-mode scheme as in Fig.~\ref{fig:twisted-mode_new}.
The twisted-mode scheme reduces the interference pattern of the four Gaussian beams shown in Fig.~\ref{fig:V-shaped_amplifier} to a two-beam interference (cf. to
  Fig.~\ref{fig:four_beams_interference}) resulting in a maximal intensity of $8~I_0$. 
The interference between $E_1$ and $E_4$ and between $E_2$ and $E_3$ is suppressed while it is unchanged between $E_1$ and $E_2$ and between $E_3$ and $E_4$.  }
\label{fig:twisted_mode_interfero}
\end{figure}

The twisted-mode scheme could be implemented also for thin-disk lasers with V-shaped cavities as shown for example in Fig.~\ref{fig:twisted-mode_new}.
Yet, contrary to the cavity of Fig.~\ref{fig:twisted-mode_classic}, in this case the twisted-mode scheme does not completely eliminate the interference.
The interference between the beams traveling to, and being back-reflected from the mirror M1 (rear side of the disk) can not be eliminated.
%
%
The resulting interference pattern which is shown in Fig.~\ref{fig:twisted_mode_interfero} reduces thus from a four-beam to a two-beam interference similar to Fig.~\ref{fig:thin-disk_end-mirror_interference}.

This reduction of the four-beam interference to a two-beam interference can be seen in the mathematical expression of the saturation spectrum $S_{SHG}\left(\lambda \right)$ which takes the form:
\begin{equation}
S_\mathrm{SHB}(\lambda)\approx 1 - 4\frac{\tau~I_0}{F_\mathrm{sat}}\left(1 + \frac{1}{2} \mathrm{sinc} \left(\pi n L \cos(\alpha) \Delta \right)\right)
\label{Eq: Twisted}
\end{equation}
similar to Eq.~(\ref{eq:end-mirror}).
This equation also represents the saturation spectrum for a unidirectional ring cavity provided $4I_0$ is substituted by $2I_0$.

Practically, this twisted-mode scheme can be implemented using only a single quarter-wave plate as shown in Fig.~\ref{fig:twisted-mode_new}.
This design is possible given the small incident angle $\alpha$ that can be realized and the availability of commercial waveplates with retardation showing only small position and incident angle dependences (the retardation scales quadratically with the incident angle) \cite{siegman1986lasers, EdmundOpticsInc, CrystechInc}.

The usage of a multi-order waveplate is advisable as it provides higher damage threshold compared to a single-order waveplate.
%
%
Beside enabling the implementation of the twisted-mode scheme the multi-order waveplate combined with a polarizer also acts as a birefringent filter \cite{Walther1970, Oka1995}.
The wavelength-dependent retardation in a round-trip that occurs in the four passes in the  quarter-wave plate leads to a wavelength-dependent round-trip transmission $B(\lambda)$
\begin{equation}
  B\left(\lambda \right) = \ln \left( 1-\cos^2{\left(2\pi~K ~\frac{\lambda -\lambda_0}{\lambda_0}\right)}\right),
  \label{eq:polarization-losses}
\end{equation}
where  $2\pi~K$ is the round-trip retardation experienced in the quarter-wave plate at the $\lambda_0$ wavelength (K = 58 for four passes in a 1.5~mm thick quartz plate at 1030~nm).
The logarithm is used to present the transmission as coefficient of the exponential in Eq.~(\ref{eq:Descrip_gain}). 

\begin{figure}[btp]
\centering
    \includegraphics[width=.5 \linewidth]{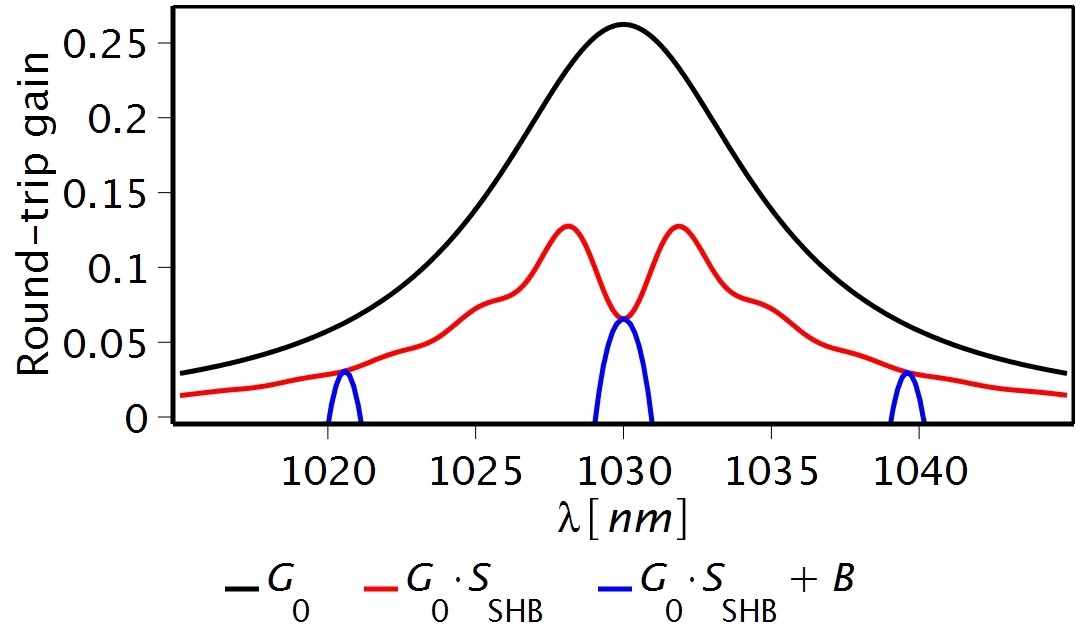} 
  \caption{Round-trip gain of the cavity depicted in Fig.~\ref{fig:twisted-mode_new} as a function of the wavelength $\lambda$. The black curve represents the unsaturated gain spectrum (we assumed a Lorentzian function as gain profile of Yb:YAG). The red curve shows the saturated gain profile for a laser operation at $\lambda_0=1030$~nm that account for SHB. We assumed $4\tau I_0 / F_\mathrm{sat}=50\%$. The blue curve represents the total round-trip gain which includes the saturated gain profile and the additional losses related to the multi-order waveplate as described by Eq.~(\ref{eq:polarization-losses}).}
\label{image15}
\end{figure}

The improved frequency selectivity achieved with the multi-order waveplate is demonstrated in Fig.~\ref{image15}.
The black curve represents a schematic representation of the unsaturated roundtrip gain $G_0$ of the cavity depicted in Fig.~\ref{fig:twisted-mode_new} using a thin disk made of Yb:YAG ($G_0(1030~\mathrm{nm}) = 0.26$ ).
%
%
The red curve 
represents the round-trip gain when accounting for the saturation effects caused by the dominant mode at $\lambda_0=1030$~nm.
It is computed by multiplying $S_\mathrm{SHB}$ from Eq.~(\ref{Eq: Twisted}) with the unsaturated gain (black curve).
Because the modes located at the two maxima of this curve see a larger gain than the dominant laser mode, they will grow until their saturation effects become relevant.
Single-frequency operation is thus disrupted.
The blue curve is the total round-trip gain that also includes the wavelength-dependent transmission $B(\lambda)$ of Eq.~(\ref{eq:polarization-losses}) for a quarter-wave plate based on a 1.5~mm thick quartz plate.
This curve shows maximal gain for $\lambda_0$. 
Hence, the twisted-mode scheme can also be used for V-shaped cavities to obtain single-frequency operation. 
%

\section{Demonstration of a Q-switched thin-disk laser operated in twisted-mode}
\label{Sec: Q-switched thin-disk laser}

For the measurement of the 2S-2P splitting in muonic helium
ions~\cite{Antognini2011, Pohl2010} we realized a
Q-switched thin-disk oscillator as sketched in Fig.~\ref{image16}.
\begin{figure}[t!]
\centering
  \begin{subfigure}[t]{.45 \linewidth}
    \centering
    \includegraphics[width= \linewidth]{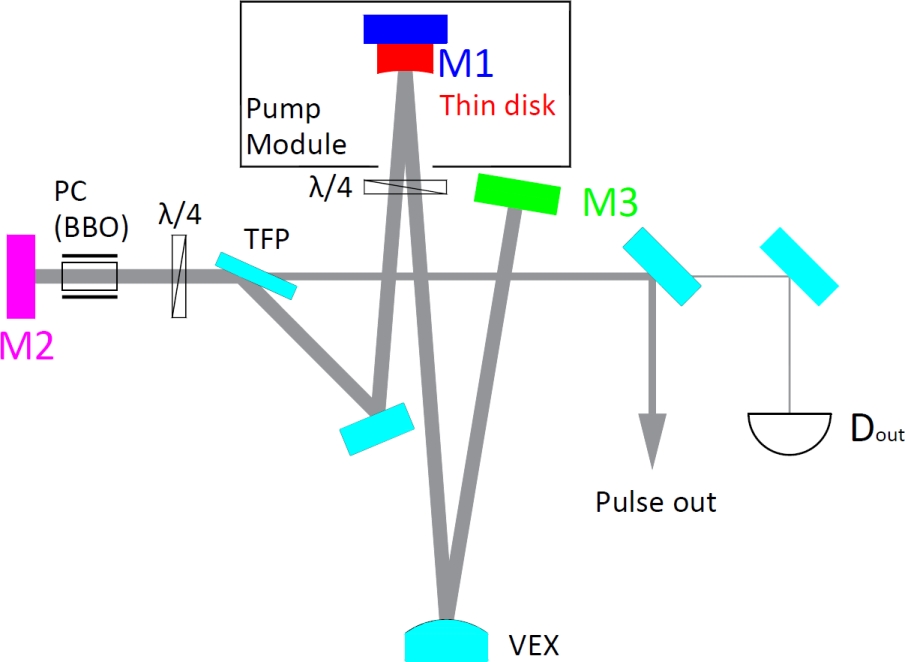} 
    \caption{} \label{image16}
  \end{subfigure}
  \hfill
  \begin{subfigure}[t]{ .45 \linewidth}
    \centering
    \includegraphics[width= \linewidth]{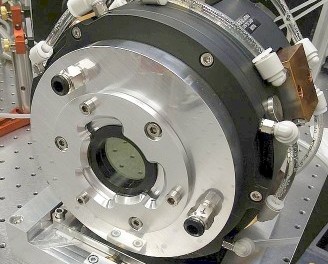} 
    \caption{} \label{image17}
  \end{subfigure}
\caption{(a): Scheme of the Q-switched thin-disk laser with twisted-mode and V-shaped layout.
M2 and M3 are the end-mirrors, TFP the thin film polarizer and PC the Pockels cell.
(b): Picture of the multi-order quarter-wave plate mounted to a TRUMPF pump module \cite{TRUMPF} acting also as window.}
\label{image16}
\end{figure}
The disk was used as a folding mirror and a multi-order quarter-wave plate was inserted in front of the disk to implement the twisted-mode scheme and to provide the additional polarization dependent losses $B(\lambda)$.
The waveplate was acting also as window for the pump module (see
Fig.~\ref{image17}) which shielded the hot front surface of the disk
from air flows that would lead to optical distortions of the laser
beam.

The other optical elements of the cavity, namely the Pockels-cell (PC), a second quarter-wave plate and a thin film polarizer (TFP) were used to control the pre-lasing operation and the Q-switch dynamics as explained for a similar system in reference~\cite{Aldo09}.
Pulses up to an energy of 110~mJ at a repetition rate of 800~Hz were delivered by the oscillator.
This laser was operated flawlessly for several months  at the high-intensity proton accelerator of the Paul Scherrer Institute, Villigen, Switzerland.

\begin{figure}[htbp]
\centering
  \begin{subfigure}[t]{.3 \linewidth}
    \centering
    \includegraphics[width= \linewidth]{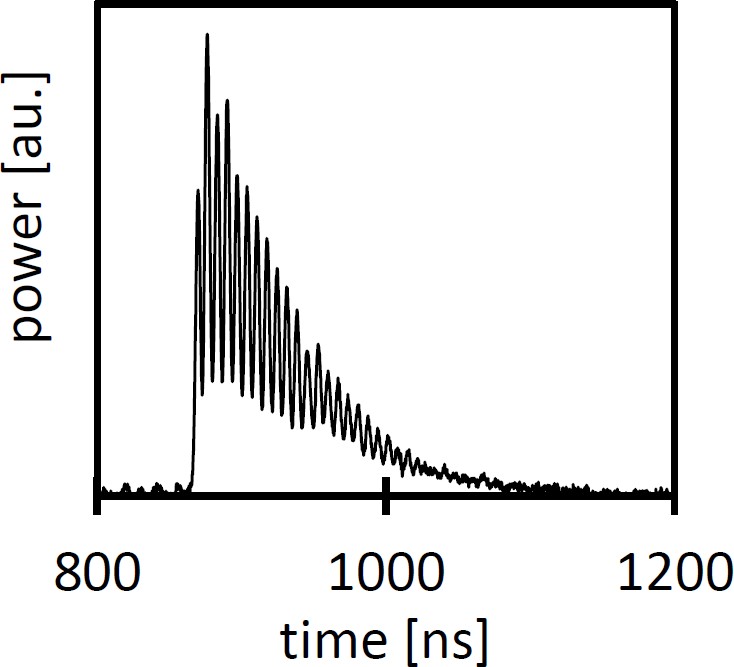} 
    \caption{} \label{Beating}
  \end{subfigure}
  \begin{subfigure}[t]{ .285\linewidth}
    \centering
    \includegraphics[width= \linewidth]{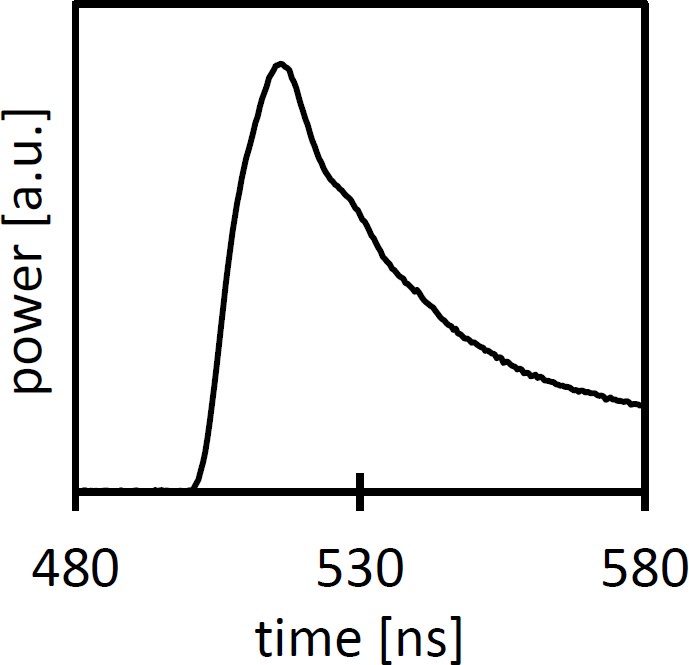} 
    \caption{} \label{Flat}
  \end{subfigure}
  \caption{Pulses emitted from the Q-switched thin-disk laser of Fig.~\ref{image16}, the time offsets are arbitrary.  In (a) the multi-order quarter-wave plate mounted in front of the disk was rotated so that the beating is maximal.  In (b) the quarter-wave plate was adjusted to produce an integer value of $K$ that results in a suppression of the beating. Here, the pulse length is slightly shorter compared to panel (a) because during this measurement the extracted energy was larger. }
\label{image18}
\end{figure}

The effectiveness of the twisted-mode scheme in reducing SHB manifests itself also in the measured temporal profile of the emitted laser pulses.
Without the quarter-wave plate, or with its axis misaligned, the SHB causes simultaneous lasing at a few longitudinal modes, resulting in mode beating with a frequency given by the round-trip time \cite{Aldo09, Berry1981, Caprara1995} (see Fig.~\ref{Beating}). 
This beating (fast oscillations) is suppressed when the quarter-wave plate is correctly aligned as demonstrated in Fig.~\ref{Flat}.
%
%
The measurement of the pulse trace is thus a sensitive method to show
the suppression of multi-mode operation~\cite{Berry1981,
  Caprara1995}.
In principle also a spectrometer could be used but a resolution
better than 50~MHz is needed.
\section{Conclusions}

In this paper we presented the wavelength-dependent decrease of the
gain in an active medium caused by saturation effects.
This decrease has been expressed as $S_\mathrm{SHB} (\lambda)$ for various linear-cavity layouts: simple cavity (Eq.~(\ref{eq:gain_HB_final}) and Fig.~\ref{fig:gain_SHB}), thin-disk laser with the thin disk acting as end-mirror (Eq.~(\ref{eq:end-mirror}) and Fig.~\ref{fig:gain-HB_end-mirror}), and thin-disk laser with the disk as folding-mirror (Eq.~(\ref{eq:G_HS_four_beams}) and Fig.~\ref{fig:G_SHB_four_waves_1}).
We found that the spectral distance between adjacent minima and maxima in $S_\mathrm{SHB}$ around the saturating wavelength is large when the disk is used as  end-mirror of the cavity. 
By contrast, it is small when the disk is used as a folding-mirror. 
Hence, the ``gain-at-the-end'' configuration allows for easy suppression of interference effects using standard selective elements as Fabry-Perot etalons. 
For the more elaborate geometry where the disks acts as folding-mirror, suppression of these maxima is more challenging given the small spectral distance (in the order of 100~MHz).

We have demonstrated that the implementation of the twisted-mode scheme
in thin-disk lasers with V-shaped layouts suppresses only partially the
SHB in the disk.
Using the twisted-mode scheme with a multi-order waveplate an additional wavelength-dependent loss arises that is spectrally sufficiently narrow to suppress the residual SHB effects that prevent single-frequency operation.
The simplicity, efficiency and robustness (in term of optical damage) of this scheme pave the way for achieving single-frequency operation of thin-disk lasers in the multi kW range.

\section*{Funding Information}

We acknowledge the support from the Swiss National Science Foundation Project SNF 200021\_165854, the European Research Council ERC CoG. \#725039 and ERC StG. \# 279765. 
The work was supported by ETH Research Grant ETH-35 14-1.
The study has been also supported by the ETH Femtosecond and Attosecond Science and Technology (ETH-FAST) initiative as part of the NCCR MUST program.

\bibliography{Thin_disk_twisted_mode}

\begin{thebibliography}{31}
\providecommand{\natexlab}[1]{#1}
\providecommand{\url}[1]{\texttt{#1}}
\expandafter\ifx\csname urlstyle\endcsname\relax
  \providecommand{\doi}[1]{doi: #1}\else
  \providecommand{\doi}{doi: \begingroup \urlstyle{rm}\Url}\fi

\bibitem[Siegman(1986)]{siegman1986lasers}
A.E. Siegman.
\newblock \emph{Lasers}.
\newblock University Science Books, 1986.
\newblock ISBN 9780935702118.
\newblock URL \url{https://books.google.ch/books?id=1BZVwUZLTkAC}.

\bibitem[Mossakowska et~al.(1993)Mossakowska, Szczepański, and
  Woliński]{Mossakowska93}
Agnieszka Mossakowska, Paweł Szczepański, and Wieslaw Woliński.
\newblock Influence of spatial hole burning effects on relaxation oscillations
  in waveguide distributed feedback {Nd$^{3+}$:YAG} lasers.
\newblock \emph{Optics Communications}, 100\penalty0 (1):\penalty0 153 -- 158,
  1993.

\bibitem[Stevens et~al.(2016)Stevens, Schlichting, Foundos, Payne, and
  Rogers]{LATJ:LATJ201600017}
Kevin~T. Stevens, Wolfgang Schlichting, Gregory Foundos, Alexis Payne, and Evan
  Rogers.
\newblock Promising materials for high power laser isolators.
\newblock \emph{Laser Technik Journal}, 13\penalty0 (3):\penalty0 18--21, 2016.
\newblock ISSN 1863-9119.
\newblock \doi{10.1002/latj.201600017}.
\newblock URL \url{http://dx.doi.org/10.1002/latj.201600017}.

\bibitem[Evtuhov and Siegman(1965)]{Evtuhov1965}
V.~Evtuhov and A.~E. Siegman.
\newblock A ``twisted-mode'' technique for obtaining axially uniform energy
  density in a laser cavity.
\newblock \emph{Appl. Opt.}, 4\penalty0 (1):\penalty0 142 -- 143, Jan 1965.

\bibitem[Kimura et~al.(1971)Kimura, Otsuka, and Saruwatari]{Kimura71}
T.~Kimura, K.~Otsuka, and M.~Saruwatari.
\newblock Spatial hole-burning effects in a {Nd$^{3+}$:YAG} laser.
\newblock \emph{IEEE Journal of Quantum Electronics}, 7\penalty0 (6):\penalty0
  225 -- 230, Jun 1971.
\newblock ISSN 0018-9197.
\newblock \doi{10.1109/JQE.1971.1076746}.

\bibitem[Smith(1972)]{Smith1972}
P.~W. Smith.
\newblock Mode selection in lasers.
\newblock \emph{Proceedings of the IEEE}, 60\penalty0 (4):\penalty0 422 -- 440,
  April 1972.
\newblock ISSN 0018-9219.
\newblock \doi{10.1109/PROC.1972.8649}.

\bibitem[Park et~al.(1984)Park, Giuliani, and Byer]{Park1984}
Y.~Park, G.~Giuliani, and R.~Byer.
\newblock Single axial mode operation of a {Q}-switched {Nd:YAG} oscillator by
  injection seeding.
\newblock \emph{IEEE Journal of Quantum Electronics}, 20\penalty0 (2):\penalty0
  117--125, February 1984.
\newblock ISSN 0018-9197.
\newblock \doi{10.1109/JQE.1984.1072371}.

\bibitem[Antognini et~al.(2009)Antognini, Schuhmann, Amaro, Biraben, Dax,
  Giesen, Graf, H\"ansch, Indelicato, Julien, Kao, Knowles, Kottmann, Bigot,
  Liu, Ludhova, Moschuring, Mulhauser, Nebel, Nez, Rabinowitz, Schwob, Taqqu,
  and Pohl]{Aldo09}
A.~Antognini, K.~Schuhmann, F.~D. Amaro, F.~Biraben, A.~Dax, A.~Giesen,
  T.~Graf, T.~W. H\"ansch, P.~Indelicato, L.~Julien, C.~Y. Kao, P.~E. Knowles,
  F.~Kottmann, E.~Le Bigot, Y.~W. Liu, L.~Ludhova, N.~Moschuring, F.~Mulhauser,
  T.~Nebel, F.~Nez, P.~Rabinowitz, C.~Schwob, D.~Taqqu, and R.~Pohl.
\newblock Thin-disk {Yb:YAG} oscillator-amplifier laser, ase, and effective
  {Yb:YAG} lifetime.
\newblock \emph{IEEE Journal of Quantum Electronics}, 45\penalty0 (8):\penalty0
  993 -- 1005, Aug 2009.

\bibitem[Giesen(2005)]{Giesen2005}
Adolf Giesen.
\newblock Thin disk lasers – power scalability and beam quality.
\newblock \emph{Laser Technik Journal}, 2\penalty0 (2):\penalty0 42 -- 45,
  2005.
\newblock ISSN 1863-9119.

\bibitem[Giesen(2007)]{Giesen2007}
Adolf Giesen.
\newblock High-power thin-disk lasers.
\newblock In \emph{Advanced Solid-State Photonics}, page MA1. Optical Society
  of America, 2007.
\newblock \doi{10.1364/ASSP.2007.MA1}.
\newblock URL
  \url{http://www.osapublishing.org/abstract.cfm?URI=ASSP-2007-MA1}.

\bibitem[Vorholt and Wittrock(2015)]{Vorholt2015}
Christian Vorholt and Ulrich Wittrock.
\newblock Spatial hole burning in {Yb:YAG} thin-disk lasers.
\newblock \emph{Applied Physics B}, 120\penalty0 (4):\penalty0 711 -- 721, Sep
  2015.

\bibitem[Paschotta et~al.(2001)Paschotta, Aus~der Au, Sp{\"u}hler, Erhard,
  Giesen, and Keller]{Paschotta2001}
R.~Paschotta, J.~Aus~der Au, G.J. Sp{\"u}hler, S.~Erhard, A.~Giesen, and
  U.~Keller.
\newblock Passive mode locking of thin-disk lasers: effects of spatial hole
  burning.
\newblock \emph{Applied Physics B}, 72\penalty0 (3):\penalty0 267--278, Feb
  2001.

\bibitem[Palmer et~al.(2008)Palmer, Schultze, Siegel, Emons, B\"{u}nting, and
  Morgner]{Palmer2008}
Guido Palmer, Marcel Schultze, Martin Siegel, Moritz Emons, Udo B\"{u}nting,
  and Uwe Morgner.
\newblock Passively mode-locked {Yb:KLu(WO$^4$)$^2$} thin-disk oscillator
  operated in the positive and negative dispersion regime.
\newblock \emph{Opt. Lett.}, 33\penalty0 (14):\penalty0 1608 -- 1610, Jul 2008.
\newblock \doi{10.1364/OL.33.001608}.
\newblock URL \url{http://ol.osa.org/abstract.cfm?URI=ol-33-14-1608}.

\bibitem[Tang et~al.(1963)Tang, Statz, and deMars]{Tang1963}
C.~L. Tang, H.~Statz, and G.~deMars.
\newblock Spectral output and spiking behavior of solid‐state lasers.
\newblock \emph{Journal of Applied Physics}, 34\penalty0 (8):\penalty0 2289 --
  2295, 1963.

\bibitem[Roess(1966)]{Roess1966}
Dieter Roess.
\newblock Analysis of a room‐temperature cw ruby laser of 10 mm resonator
  length: The ruby laser as a thermal lens.
\newblock \emph{Journal of Applied Physics}, 37\penalty0 (9):\penalty0 3587 --
  3594, 1966.

\bibitem[Baer(1986)]{Baer1986}
T.~Baer.
\newblock Large-amplitude fluctuations due to longitudinal mode coupling in
  diode-pumped intracavity-doubled {Nd:YAG} lasers.
\newblock \emph{J. Opt. Soc. Am. B}, 3\penalty0 (9):\penalty0 1175 -- 1180, Sep
  1986.

\bibitem[Schuhmann et~al.(2013)Schuhmann, Antognini, Kirch, Graf, Ahmed, Voss,
  and Weichelt]{Schuhmann2013}
Karsten Schuhmann, Aldo Antognini, Klaus Kirch, Thomas Graf, Marwan~Abdou
  Ahmed, Andreas Voss, and Birgit Weichelt.
\newblock Thin-disk laser for the measurement of the radii of the proton and
  the alpha-particle.
\newblock In \emph{Advanced Solid-State Lasers Congress}, page ATu3A.46.
  Optical Society of America, 2013.

\bibitem[Walther and Hall(1970)]{Walther1970}
H.~Walther and J.~L. Hall.
\newblock Tunable dye laser with narrow spectral output.
\newblock \emph{Applied Physics Letters}, 17\penalty0 (6):\penalty0 239 -- 242,
  1970.
\newblock \doi{10.1063/1.1653382}.
\newblock URL \url{http://dx.doi.org/10.1063/1.1653382}.

\bibitem[Magni(1986)]{Magni1986}
Vittorio Magni.
\newblock Resonators for solid-state lasers with large-volume fundamental mode
  and high alignment stability.
\newblock \emph{Appl. Opt.}, 25\penalty0 (1):\penalty0 107--117, Jan 1986.
\newblock \doi{10.1364/AO.25.000107}.
\newblock URL \url{http://ao.osa.org/abstract.cfm?URI=ao-25-1-107}.

\bibitem[Bollen et~al.(1987)Bollen, Kluge, Wallmeroth, Schaaf, and
  Moore]{Bollen1987}
G.~Bollen, H.-J. Kluge, K.~Wallmeroth, H.~W. Schaaf, and R.~B. Moore.
\newblock High-power pulsed dye laser with {Fourier-limited} bandwidth.
\newblock \emph{J. Opt. Soc. Am. B}, 4\penalty0 (3):\penalty0 329 -- 336, Mar
  1987.

\bibitem[Adams et~al.(1993)Adams, Vorberg, and Mlynek]{Adams1993}
C.~S. Adams, J.~Vorberg, and J.~Mlynek.
\newblock Single-frequency operation of a diode-pumped
  lanthanum-neodymium-hexaaluminate laser by using a twisted-mode cavity.
\newblock \emph{Opt. Lett.}, 18\penalty0 (6):\penalty0 420 -- 422, Mar 1993.
\newblock \doi{10.1364/OL.18.000420}.
\newblock URL \url{http://ol.osa.org/abstract.cfm?URI=ol-18-6-420}.

\bibitem[Pan et~al.(2005)Pan, Xu, and Zeng]{Pan2005}
Haifeng Pan, Shixiang Xu, and Heping Zeng.
\newblock Passively {Q}-switched single-longitudinal-mode c-cut {Nd:GdVO$^4$}
  laser with a twisted-mode cavity.
\newblock \emph{Opt. Express}, 13\penalty0 (7):\penalty0 2755 -- 2760, Apr
  2005.
\newblock \doi{10.1364/OPEX.13.002755}.
\newblock URL \url{http://www.opticsexpress.org/abstract.cfm?URI=oe-13-7-2755}.

\bibitem[Cong et~al.(2016)Cong, Liu, Zhang, Qin, Men, Fu, and Rao]{Cong2016}
Z.~H. Cong, Z.~J. Liu, X.~Y. Zhang, Z.~G. Qin, S.~J. Men, Q.~Fu, and H.~Rao.
\newblock Actively {Q}-switched {N}d:{YAG} twisted-mode cavity laser with a
  {RTP} electro-optic modulator.
\newblock In \emph{2016 International Conference Laser Optics (LO)}, pages
  R1--65--R1--65, June 2016.

\bibitem[{Edmund Optics Inc.}(2017)]{EdmundOpticsInc}
{Edmund Optics Inc.}
\newblock Understanding waveplates, 06 2017.
\newblock {h}ttps://www.edmundoptics.com/resources/application-notes/optics/
  understanding-waveplates/.

\bibitem[{Crystech Inc.}(2017)]{CrystechInc}
{Crystech Inc.}
\newblock Waveplates, 06 2017.
\newblock
  {h}ttp://ilphotonics.com/cdv2/Crystech-Crystals-Optics/Optics-Crystech/waveplate.pdf/.

\bibitem[Oka et~al.(1995)Oka, Liu, Wiechmann, Eguchi, and Kubota]{Oka1995}
M.~Oka, Ling~Yi Liu, W.~Wiechmann, N.~Eguchi, and S.~Kubota.
\newblock All solid-state continuous-wave frequency-quadrupled {N}d:{YAG}
  laser.
\newblock \emph{IEEE Journal of Selected Topics in Quantum Electronics},
  1\penalty0 (3):\penalty0 859 -- 866, Sep 1995.
\newblock ISSN 1077-260X.
\newblock \doi{10.1109/2944.473671}.

\bibitem[Antognini et~al.(2011)Antognini, Biraben, Cardoso, Covita, Dax,
  Fernandes, Gouvea, Graf, Hänsch, Hildebrandt, Indelicato, Julien, Kirch,
  Kottmann, Liu, Monteiro, Mulhauser, Nebel, Nez, dos Santos, Schuhmann, Taqqu,
  Veloso, Voss, and Pohl]{Antognini2011}
A.~Antognini, F.~Biraben, J.~M.R. Cardoso, D.~S. Covita, A.~Dax, L.~M.P.
  Fernandes, A.~L. Gouvea, T.~Graf, T.~W. Hänsch, M.~Hildebrandt,
  P.~Indelicato, L.~Julien, K.~Kirch, F.~Kottmann, Y.-W. Liu, C.~M.B. Monteiro,
  F.~Mulhauser, T.~Nebel, F.~Nez, J.~M.F. dos Santos, K.~Schuhmann, D.~Taqqu,
  J.~F.C.A. Veloso, A.~Voss, and R.~Pohl.
\newblock Illuminating the proton radius conundrum: the $\mu${He}$^+$ lamb
  shift.
\newblock \emph{Canadian Journal of Physics}, 89\penalty0 (1):\penalty0 47 --
  57, 2011.

\bibitem[Pohl et~al.(2010)Pohl, Antognini, Nez, Amaro, Biraben, Cardoso,
  Covita, Dax, Dhawan, Fernandes, Giesen, Graf, H\"ansch, Indelicato, Julien,
  Kao, Knowles, Le~Bigot, Liu, Lopes, Ludhova, Monteiro, Mulhauser, Nebel,
  Rabinowitz, dos Santos, Schaller, Schuhmann, Schwob, Taqqu, Veloso, and
  Kottmann]{Pohl2010}
Randolf Pohl, Aldo Antognini, François Nez, Fernando~D. Amaro, François
  Biraben, Jo\~{a}o M.~R. Cardoso, Daniel~S. Covita, Andreas Dax, Satish
  Dhawan, Luis M.~P. Fernandes, Adolf Giesen, Thomas Graf, Theodor~W. H\"ansch,
  Paul Indelicato, Lucile Julien, Cheng-Yang Kao, Paul Knowles, Eric-Olivier
  Le~Bigot, Yi-Wei Liu, Jos\'{e} A.~M. Lopes, Livia Ludhova, Cristina M.~B.
  Monteiro, Fran\c{c}oise Mulhauser, Tobias Nebel, Paul Rabinowitz, Joaquim
  M.~F. dos Santos, Lukas~A. Schaller, Karsten Schuhmann, Catherine Schwob,
  David Taqqu, Jo\~{a}o F. C.~A. Veloso, and Franz Kottmann.
\newblock The size of the proton.
\newblock \emph{Nature}, 466:\penalty0 213 --– 216, July 2010.

\bibitem[{TRUMPF GmbH}(2017)]{TRUMPF}
{TRUMPF GmbH}.
\newblock Scheibenlaser, 06 2017.
\newblock
  {h}ttp://www.trumpflaser.com/de/produkte/festkoerperlaser/scheibenlaser
  /trudisk.htm.

\bibitem[Berry et~al.(1981)Berry, Hanna, and Sawyers]{Berry1981}
A.J. Berry, D.C. Hanna, and C.G. Sawyers.
\newblock {High power, single frequency operation of a {Q}-switched {TEM00}
  Mode {Nd:YAG} laser}.
\newblock \emph{Optics Communications}, 40\penalty0 (1):\penalty0 54 -- 58,
  1981.
\newblock ISSN 0030-4018.
\newblock \doi{http://dx.doi.org/10.1016/0030-4018(81)90270-4}.
\newblock URL
  \url{http://www.sciencedirect.com/science/article/pii/0030401881902704}.

\bibitem[Caprara and Heritier(1995)]{Caprara1995}
A.L. Caprara and J.M. Heritier.
\newblock Single longitudinal mode laser without seeding, May~2 1995.
\newblock URL \url{https://www.google.com/patents/US5412673}.
\newblock US Patent 5,412,673.

\end{thebibliography}
\bibliographystyle{plain}




\end{document}